\documentclass[ preprint]{aastex}
\usepackage{epsfig,rotate}
\usepackage{graphicx}
\newcommand{\be}{\begin{equation}}
\newcommand{\ee}{\end{equation}}

\def\be{\begin{equation}}
\def\ee{\end{equation}}
\def\beq{\begin{equation}}
\def\eeq{\end{equation}}

\def\etal{~{\it et al.\ }}
\def\bea{\begin{eqnarray}}
\def\eea{\end{eqnarray}}
\def\noi{\noindent}

\begin{document}

\title{Resonantly-forced Eccentric Ringlets: Relationships between Surface
Density, Resonance location, Eccentricity and Eccentricity-gradient}

\author{M.D. Melita}
\author{J.C.B. Papaloizou}
\affil{Astronomy Unit,\\  Queen Mary, University of London,\\
Mile End Rd.,\\  London, E14NS.}



\begin{abstract}

We use a simple model of the dynamics of a narrow-eccentric ring, to put
some constraints on some of the observable properties of the real systems.
In this work we concentrate on the case of the `Titan ringlet of Saturn'.

Our approach is fluid-like, since our description is based on normal modes
of oscillation rather than in individual particle orbits. Thus, the rigid
precession of the ring is described as a global $m=1$ mode, which originates
from a standing wave superposed on an axisymmetric background. An integral
balance condition for the maintenance of the $m=1$ standing-wave can be set
up, in which the differential precession induced by the oblateness of the
central planet must cancel  the contributions of self-gravity,
the resonant satellite forcing and collisional effects.  We expect that
in nearly-circular narrow rings dominated by self-gravity, the
eccentricity varies linearly across the ring. Thus, we take a first order
expansion and we derive two integral relationships from the rigid-precession
condition. These relate the surface density of the ring with the
eccentricity at the center, the eccentricity gradient and the location of
the secular resonance. 

These relationships are applied to the Titan ringlet of Saturn, which has a
secular resonance with the satellite Titan in which the ring precession
period is close to Titan's orbital period. In this case, we  estimate
the mean surface density and the location of the secular resonance.  

\end{abstract}
%
%

\section{Introduction}

The dynamical mechanism that maintains the apse alignment of the observed
narrow-eccentric planetary rings is basically governed by self-gravity
(Goldreich and Tremaine 1979), which would provide the appropriate
contribution to counter-act the differential precession induced by the
oblateness of the central planet. However, predictions of the total mass of
the ring produced by this model are, in general, not in good agreement
with the inferred mass of observed eccentric rings
(Tyler\etal 1986, Graps \etal 1995, Goldreich and Porco 1987). This led to
the consideration of other factors that might play an important role in the
dynamics. In particular, at their narrowest point, the ring particles are
`close-packed'.  In such a situation particle interaction or pressure
effects may affect the precession of particle orbits. A simple model where
the {\it pinch} locks the differential precession, was introduced by Dermott
and Murray (1980). A more global picture, including the effect of stresses
due to particle interactions and neighboring satellite perturbations, which
offered a better agreement with the observations, has also been produced by
Borderies\etal (1983). Their dynamical model is described in terms of
mutually interacting {\it streamlines} and the satellite interactions (see
Goldreich \& Tremaine 1981) are computed using a resonance-continuum
approximation. The standard self-gravity model was later revisited by Chiang
and Goldreich (2000), who considered the effects of collisions near the
edges, proposing that a sharp increase of an order of magnitude in the
surface density should be observed within the last few hundred meters of the
ring edges. More recently, employing a pressure term that describes
close-packing, Mosqueira and Estrada (2002) obtained surface-density
solutions that agree well with the currently available mass estimates.

The eccentric precessing-pattern of the ring can be described as being
generated by a normal mode of oscillation of wave-number $m=1$, which can be
viewed as a standing wave. The conditions for the maintenance of steady
global $m=1$ modes have been considered by Papaloizou \& Melita (2004).  To
describe the ring perturbations and the $m=1$ mode we used the
Lagrangian-displacement of the particle orbits from their unperturbed
circular ones (see for example Shu\etal 1985). This model includes the
dissipation due to inter-particle collisions, which would lead to damping of
the mode. However, this global $m=1$ mode can also be perturbed by
neighboring-shepherd satellites, which can inject energy and angular
momentum through resonances. In this way, losses due to particle collisions
can be balanced. Two conditions for the maintenance of the rings can be
derived. The first one is a condition for the steady maintenance of the
amplitude or eccentricity associated with the $m=1$ mode, which requires the
external satellite torque to balance the dissipative effects due to
collisions (see Papaloizou \& Melita 2004). The second one is the condition
of uniform precession of the ring, which, in the absence of satellite
resonances with the $m=1$ mode,  only involves self-gravity and the effect of 
collisions. These conditions can be regarded as continuum forms of
the discrete relationships that can be obtained from the `many streamlines'
model (Goldreich and Tremaine 1979, Longaretti and Rapapport 1995). 

In this work we produce an extension of Papaloizou and Melita (2004) to the
case where the pattern frequency of the narrow-eccentric ringlet is in a
secular resonance with an external satellite. In this case, there is a
contribution to the condition of uniform precession, arising from the
resonant secular perturbation. Resonantly-forced rings are of particular
interest, because there is a real system that the model can be applied to.
It is known that the precession frequency of the eccentric Saturnian ringlet
at $1.29\ R_S$ (popularly known as the `Titan ringlet') is in a $1:0$
resonance with the orbital frequency of the Saturnian satellite Titan (Porco
et al. 1984). 

From the rigid precession condition two useful relations are derived. If the
eccentricity gradient is approximately constant across the ring, which is expected to be
the case for a narrow ringlet dominated by self-gravity, then, these
relationships can   constrain the mean surface density, the central
eccentricity, the eccentricity gradient and the location of the secular
resonance. We estimate the mean surface density as a function of the
location of the resonance and the form of the ring in the case where physical
collisions are neglected. 

This article is organized as follows. In section~\ref{LD} we set up the
equations for the Lagrangian variations starting from the equations of
motion in a 2D flat disk approximation. In section~\ref{m1m} we give 
adequate approximations for the evolution of the $m=1$ mode  when 
 the precession time is much longer than the orbital period. In
section~\ref{UP} we derive the condition of rigid precession which
incorporates secular satellite resonances and we also compute the
contribution from the self gravity of the ring. Two integral relationships
are obtained form the rigid-precession condition in section~\ref{vq}.  The
relation between the eccentricity gradient, the value of the central
eccentricity, the location of the resonance and the surface density is given
in section~\ref{qe}, where we also discuss how to obtain the eccentricity
gradient from the other parameters in the linear and the non-linear regimes.
In section~\ref{Titan} we apply these results to the Titan ringlet of
Saturn to produce estimates of its mean surface-density. Finally, a
discussion of the results is given in section~\ref{discu}.

\section{Equations of Motion and Lagrangian Displacement}
\label{LD}

We start from the basic equations of motion for a particle in Lagrangian
form in 2D:
\be 
{d^2 r\over dt^2} - r\left( {d\theta \over dt}\right)^2 = 
F_r -{\partial \psi \over \partial r}
\label{Mr}
\ee
\be 
r\ {d^2 \theta \over dt^2} + 2\left({d r \over dt}\right)\left( {d\theta
\over dt}\right) = 
F_{\theta} - {1\over r} {\partial \psi \over \partial \theta } 
\label{MTH}
\ee
\noi Here $(r,\theta)$ define the cylindrical coordinates of the
particle referred to an origin at the centre of mass of the planet. Here
$\psi(r)$ denotes the gravitational potential due to both the central
planet, the neighboring satellites and the ring. In addition $(F_r,
F_{\theta})$ denote the radial and azimuthal components of any additional
force ${\bf F}$ per unit mass respectively.  This may arise through internal
interactions between particles that might lead to an effective pressure
and/or viscosity. 

We introduce a Lagrangian description in which the system is supposed to be
perturbed from an axisymmetric state in which particles are in circular
motion with coordinates such that $r = r_0 ,$ $ \theta = \theta_0=
\Omega(r_0)t + \beta_0.$ Here $r_0$ is the fixed radius of the particle
concerned, $\Omega(r_0)$ is the angular velocity and $\beta_0$ is a phase
factor labeling each particle. In keeping with a Lagrangian description
$(r_0, \beta_0 )$ are conserved quantities for a particular particle and so
may be used to label it.

In order to describe the system when it is perturbed from the axisymmetric
state we introduce the components of the Lagrangian displacement $
{\mbox{\boldmath $\xi$ }} = (\xi_r, \xi_{\theta}).$ These are such that the
coordinates of each particle satisfy:
\be r = r_0 + \xi_r,\ee and \be r_0(\theta -\theta_0 ) = \xi_{\theta}.\ee

To obtain equations for $\xi_r$ and $\xi_{\theta}$ we take variations of
Eq.'s~({\ref{Mr}) and (\ref{MTH}). We do this by applying the Lagrangian
difference operator, $\Delta$, as defined by Lebovitz (1961) to both sides
of Eq.'s~(\ref{Mr}) and~(\ref{MTH}). For a given quantity $Q$, the
variation $\Delta(Q)$ is defined by: \be \Delta(Q) = Q\left( {\bf r}_0 + {\bf
\xi} \right) - Q_0\left( {\bf r}_0 \right), \label{Lagrdef} \ee where $Q$
and $Q_0$ are the values of the given physical quantity in the perturbed and
unperturbed flow respectively.  In contrast, the Eulerian difference
operator is defined as: \be \delta(Q) = Q\left( {\bf r}_0 \right) - Q_0\left(
{\bf r}_0 \right). \ee Thus, to first order, they are related through: \be
\Delta =\delta + {\mbox{\boldmath$\xi$}} \ .\ \nabla \ee which gives the
linear form of the Lagrangian difference operator.

\subsection {Equations for the Lagrangian displacement} Following Shu\etal
(1985) we assume that the components of the displacement are small enough
that they can be treated as linear in the sense that
$|{\mbox{\boldmath$\xi$}}/r_0| << 1.$ On the other hand the radial gradient
of the radial displacement may be large so that $|(\partial \xi_r/\partial
r_0)|$ may be of order unity. The significance of these assumptions is that
although the ring eccentricity is assumed to be everywhere small, the ring
surface density perturbation induced by it may be of order unity. Adopting
them enables us to perform the variation in the accelerations using the
linear form of the difference operator as described above, wherever radial
gradients are not involved. These then satisfy: \be {d^2 \xi_r\over dt^2}
-2\Omega {d \xi_{\theta} \over dt}+ 2\xi_r r_0 \Omega {d \Omega \over dr_0}
= f_{r} - {\partial \psi'\over \partial r} 
\label{pr}\ee \be {d^2 \xi_\theta \over dt^2} + 2\Omega {d \xi_r \over dt} =
f_{\theta} -  {1\over r} {\partial \psi' \over \partial \theta
} . \label{pth0} \ee 

Here the potential due to the satellite, $\psi_{s}$, and
that due to the self-gravity of the ring, $\psi_{SG}$, are included in
$\psi'.$ Thus $\psi' = \psi_{SG} + \psi{s}.$ The quantities $f_r =
\Delta(F_{r}),\ f_\theta = \Delta(F_{\theta})$ denote the variational
components of the force per unit mass due to particle interactions. The full
non linear Lagrangian variation is retained for $\psi'$ and ${\bf F}$ as
these may involve the density variation. Contributions coming from the
variation of the central planet potential are included on the left hand sides
of Eq.~(\ref{pr}) and Eq.~(\ref{pth0}).

\subsection{Surface density perturbation and Lagrangian time derivatives} 

We suppose the ring particles to be in eccentric orbits and combine to form
a globally eccentric ring. This is described using a surface density
distribution $\Sigma( r,\theta )$ and eccentricity distribution $e(r).$ We
also consider there to be an axisymmetric reference state for which $e(r) =
\xi_r /r_0 $ and from which we can regard the eccentric ring as being the
result of a perturbation. The perturbation of the surface density is of the
form:
\be \Sigma( r, \theta, t)\rightarrow \Sigma( r, \theta ) + \Sigma'( r,
\theta, t).
\ee
For linear perturbations $\Sigma' \propto \cos / \sin (m\theta),$ where the
azimuthal mode number, $m = 1.$ The eccentric ring can be thought of as being
predominantly in a mode with azimuthal mode number $m=1.$ In practice we may
assume $|e| << 1.$

\noindent  We further remark that the convective derivative $d \ /dt$
is taken  following the  fluid motion. In the
approximation scheme used here in which the displacements
and hence Lagrangian velocity perturbations are small,
we may replace the fluid motion by its unperturbed value.
Then for any quantity $Q$ 
\be
{dQ \over dt} = {\partial Q \over \partial t} + \Omega  {\partial Q
\over \partial \theta_0} . \label{convd} \ee 
Similarly
\be 
{d^2 Q \over dt^2} = {\partial^2 Q \over \partial t^2} +\Omega^2 
 {\partial^2 Q
\over \partial \theta_0 ^2}
+ 2\Omega {\partial^2 Q
\over {\partial \theta_0 \partial t }} . \label{convd2}
\ee

\section{Forcing of the m=1 (eccentric) mode}
\label{m1m}

\noindent We here consider the forcing of the
$m=1$ mode which causes the ring to become eccentric.
The forcing is assumed to be due to an external satellite
with mass $M_S.$
With an aim of application to the Titan ringlet around Saturn
we consider that the perturbing potential acting on the ring
is stationary in a frame rotating with the mean  orbital
rotation rate when viewed from
an inertial frame.   In terms of the ring dynamics this pattern rotates at a low
frequency, $\Omega_p,$  
 such that 
$\Omega_p << \Omega.$
A free $m = 1$ mode  that is most easily excited is one that
has a global structure in the ring and has a  pattern that precesses
 at a rate comparable to $\Omega_p.$ When this precession rate
is equal to $\Omega_p.$ there is the  possibility
of resonance and a large response to forcing.
The quantity $\Omega_p^{-1}$ sets the natural time scale
for variations associated with the $m=1$ mode for the problem on hand.
 
\noindent Accordingly: \be {\partial \over \partial t} \ll \Omega \left(
{\partial \over \partial \theta_0}\right ) \label{ineq}\ee Recalling that
the left hand side of Eq.~(\ref{pth0}) approximated by the linearized
form, gives for the azimuthal component of the displacement:
\be {d \xi_\theta \over dt} + 2\Omega \xi_r   = Q_{\theta_0} \label{qth}\ee

\noindent where the quantity $Q_{\theta_0}$ is defined by:
\be {\partial Q_{\theta_0} \over \partial t} + \Omega {\partial Q_{\theta_0}
\over \partial \theta_0}= f_{\theta} - {1\over r} {\partial
\psi' \over \partial \theta_0 } . \label{pth}\ee

Using  (\ref{ineq}) gives the adequate approximation:
\be \Omega {\partial Q_{\theta_0} \over \partial \theta_0}= f_{\theta} -
 {1\over r} {\partial \psi' \over \partial \theta_0 }.
\label{poth}\ee

\noindent We comment that the motion is dominated by the central mass and to the
lowest order Keplerian, This means that the $m=1$ component of the
displacement satisfies (Shu\etal 1985): 
\be {\partial^2 \xi_r\over \partial \theta_0^2 } = - \xi_r. \ee

\noindent Furthermore ({\ref{qth}) tells us 
to lowest order in which $Q_{\theta_0}$
and ${\partial \over \partial t}$ may be neglected that: \be { \partial
\xi_\theta \over \partial \theta_0 } = - 2 \xi_r \label{Kep}\ee which
applies to Keplerian orbits with small eccentricity.

\noindent Also using (\ref{qth})  
and  Eq.~(\ref{pr}) one
finds that the $m=1$ component of the displacement satisfies:
\be {\partial^2 \xi_r \over \partial t^2} + 
2\Omega {\partial^2 \xi_r\over \partial t \partial \theta_0} - \xi_r
(\Omega^2 - \kappa^2 )= f_{r} -  {\partial \psi' \over
\partial r} + 2\Omega Q_{\theta_0},\label{rmot}\ee

\noindent Here the square of the epicyclic frequency is given by: \be \kappa^2
={2\Omega \over r_0}{d (r_0^2 \Omega)\over dr_0}. \label{epi}\ee

\section{The condition for a steady state response in the rotating frame}
\label{UP}

\noindent The $m=1$ mode responsible for the ring eccentricity has a constant
and very small pattern speed as viewed in the inertial frame. This
means that individual ring particles appear to be in elliptic orbits
that precess at the same rate. In oder to achieve this the internal
and external forces acting in the mode have to satisfy a constraint
that can be view as a non-linear dispersion relation. Our treatment
again follows that of Shu\etal (1985) who provided such a relationship
for density waves in Saturn's rings. Except here we consider a density
wave comprising a global normal mode rather than a forced propagating
wave.

Eq.~(\ref{rmot}) can  also be written  in the form:
\be 
{d^2 \xi_r\over dt^2} + \xi_r \kappa^2 = f_{r} - {\partial
\psi' \over \partial r}  + 2 \Omega Q_{\theta_0} \label{pr2}
\ee 
We now use an angle that is fixed with respect to a coordinate system
rotating at the pattern angular frequency $\Omega_P,$ namely $\phi_0 =
\theta_0 - \Omega_P\ t.$ The radial displacement is taken to be of the form
$\xi_r = A(r_0)\ \cos(\phi_0).$ Following Shu\etal (1985) we note that as the
time dependence is contained within $\phi_0,$ $\xi_r$ only depends on $r_0$
and $\phi_0.$
       
\noindent Multiplying Eq.~(\ref{pr2}) by $\cos(\phi_0)$ and integrating over
$\phi_0,$ we obtain: 
$$\frac{1}{2}  \left( \frac{\kappa^2}{(\Omega - \Omega_P)^2} - 1
\right)\ A(r_0) = \frac{1}{(\Omega - \Omega_P)^2} ( F_{cr} +
g_D(r_0) $$
\be  + g_R(r_0)\\
+  \frac{1}{2\pi} \int_0^{2\pi} 2\ \Omega Q_{\theta0} \cos(\phi_0) d\phi_0)
, \label{pr3} \ee where: \be F_{cr} = \frac{1}{2\pi} \int_0^{2\pi}
f_r \cos(\phi_0) d\phi_0, \ee  \be g_{D}(r_0) = - \frac{1}{2\pi}
\int_0^{2\pi} \cos(\phi_0) {\partial \psi_{SG} \over \partial
r}  d\phi_0 
\ee  and 
\be g_{R}(r_0) = - \frac{1}{2\pi}
\int_0^{2\pi} \cos(\phi_0) {\partial \psi_{s} \over \partial
r}  d\phi_0
\ee
gives the forcing due to the  satellite potential which is
here considered to be responsible for the excitation of  
the $m=1$ mode.

With the use of equation~(\ref{poth}), the last term in Eq.~(\ref{pr3}) can be
re-written and after an integration it reads as: 
\be \frac{1}{2\pi}
\int_0^{2\pi} 2\ \Omega Q_\theta \cos(\phi_0) d\phi_0 = -2\ (F_{c\theta} +
g_T) , \ee 
where: 
\be F_{c\theta} = \frac{1}{2\pi} \int_0^{2\pi} f_\theta
\sin(\phi_0) d\phi_0, \ee 
and 
\be g_T= -\frac{1}{2\pi} \int_0^{2\pi}
\sin\phi_0 {1\over r} {\partial \psi_s \over \partial \phi_0} d\phi_0 .
\ee 
Eq.~(\ref{pr3}) then becomes: 
\be \frac{1}{2} \left( \frac{\kappa^2}{(\Omega
- \Omega_P)^2} - 1 \right)\ A(r_0) = \frac{g_{int} + g_S }{(\Omega -
\Omega_P)^2}\, 
\label{pr4} 
\ee 
where 
\be g_{int} = \left( F_{cr} - 2\ F_{c\theta} \right) + g_D 
\ee 
and 
\be g_S = g_R - 2\ g_T .  
\label{gext} 
\ee 
Given that $\kappa = \Omega - \omega_{prec}$, where $\omega_{prec}(r_0)$ is
the local radius dependent precession frequency and assuming that $\Omega_P
<< \Omega$ and $\omega_{prec} << \Omega$, Eq.~(\ref{pr4}) can be
approximated to first order in $\Omega_P$ and $\omega_{prec}$ as: \be 
(\Omega_P - \omega_{prec})\ A(r_0) = \frac{g_{int} + g_S}{\Omega}
\label{pr5} \ee Eq.~(\ref{pr5}) provides a condition to be satisfied by the
excited $ m=1$ mode amplitude It balances the ring self gravity, internal
collisional terms and satellite forcing. For simplicity we shall neglect
collisional effects below and thus replace $g_{int}$ by $g_{D}.$ Note
further that for a thin ring of the type considered here, $\Omega$ may be
taken as constant in (\ref{pr5}) and evaluated at the ring centre from now
on.

\subsection{The self-gravity term}
\label{SG}

In order to calculate $g_D$ we follow Shu et al. (1985). As radial variations are
much more rapid than azimuthal ones, the local self-gravity at $r_0$ is 
canonically and adequately approximated to be that due to an infinite plane sheet of radial
width $\Delta r = r_2 - r_1$, where $r_1$ and $r_2$ are the inner and outer
bounding radii of the unperturbed ring respectively. Thus: 
\be 
\left({\partial \psi_{SG} \over \partial r} \right) = 2G\ \int_{r_1}^{r_2}
\frac{\Sigma(r')}{ (r-r')} dr' \label{gs1} 
\ee 
where $G$ is the gravitational constant, $r = r_0 + \xi_r$ and $r' = r_0' +
\xi_r'$, where 
\be 
\xi_r = \xi_r(r_0) = A(r_0)\ cos(\phi_0) 
\label{xi_r}
\ee 
and 
\[ \xi_r' = \xi_r(r_0') = A(r_0')\ cos(\phi_0)\].  

Possible singularities
in the integrand are dealt with by evaluating the integral in a principal
value sense. In the planar limit, we identify the
ring eccentricity as $e(r_0) = 2A(r_0)/(r_1+r_2).$ Using the tight-winding
approximation we have: 
\be 
\Sigma(r')dr' = \Sigma(r_0')dr_0' 
\ee 
which represents conservation of mass.  Since $r_0/r \approx 1 $, we have: 
\be 
\left({\partial \psi_{SG} \over \partial r} \right) = 2G\
\int_{r_1}^{r_2} \frac{\Sigma(r_0')}{ r_0 + \xi_r - r_0' - \xi_r'} dr_0'
\label{gs2} 
\ee 
We can re-write Eq.~(\ref{gs2}) in terms of the eccentricity gradient,
$q$:
\be q= \frac{A(r_0) - A(r_0')}{r_0 - r_0'}. 
\label{q} 
\ee 

Then after integrating over $\phi,$ we obtain (see also Shu\etal 1985): 
\be g_{D} = 2G\ \int_{r_1}^{r_2} \frac{I(q)}{q} \Sigma(r_0')\ \frac{A(r_0) -
A(r_0')}{(r_0 - r_0')^2} dr_0' \label{gs21} 
\ee 
where: 
\be 
I(q) = \frac{1}{2\pi} \int_0^{2\pi} \frac{cos(\phi)}{ 1 - q\cos{\phi}}
d\phi = \frac{1}{q \sqrt{1 - q^2}}\ \left(1 - \sqrt{1 - q^2}\right) 
\ee
Notice that the integrand in Eq.~(\ref{gs21}) presents a singularity
to be handled in a principal value sense.
This can lead to practical complications near  ring edges.

\section{Two integral relations} \label{vq} The practical problem is to
solve equation~(\ref{pr5}) for the response to the forcing by the external
satellite. This is equivalent to calculating the forced eccentricity. To do
this requires an accurate specification of the ring surface density profile
which may not be available. Instead one may derive two integral relations
which contain complete in formation about the response when it is a linear
function of radius. That is equivalent to assuming the constancy of $q$
defined above which is a frequently adopted approximation in planetary ring
dynamics (eg. Goldreich and Tremaine 1979, Borderies et al. 1983, Shu et al.
1985, Chiang and Goldreich 2000).

\noindent A justification for this is that normally (as here) one considers
the case when ring self-gravity is strong enough to balance
differential precession. In that case the ring precesses at a uniform
rate similar to 
a rigid body. 
When this process is effective, strong self-gravity
precludes short wavelength displacements so that approximating   
the induced displacement
as a linear function of the distance to the ring centre is reasonable
as long as the eccentricity response is not too large.

\noindent For example, a strict resonance between the pattern speed and the
uniform precession frequency of the ring might result in large eccentricities
being excited for which the dependence of the precession frequency on
eccentricity should not be neglected.  We note that if $e^2 > |(\Delta a/
\omega_{prec}) \times (d\omega_{prec}/dr)|$, $\Delta a$ being the ring
semi-major axis width, that effect becomes comparable to that due to
differential precession and should not be neglected --as we have done-- in
comparison to that. However, this situation is not encountered for the
application considered here.

\subsection{The first relation}
To obtain this we take equation~(\ref{pr5})
multiply by $\Sigma(r_0)$  and integrate over the ring
to obtain

\be
\int \Sigma(r_0)\Omega
(\Omega_P - \omega_{prec})\ A(r_0) dr_0
 - \int \Sigma(r_0)g_{D} dr_0 =  \int \Sigma(r_0)g_{S} dr_0 
\label{pr5r1}
\ee
From equation~(\ref{gs21}) the second integral on the left hand side
of equation~(\ref{pr5r1}) is zero.
The first relation thus simplifies to become
\be
\int \Sigma(r_0)\Omega
(\Omega_P - \omega_{prec})\ A(r_0) dr_0
 =  \int \Sigma(r_0)g_{S} dr_0
\label{pr5r11}
\ee

\subsection{The Second relation}
To obtain this we take equation~(\ref{pr5})
multiply by $\Sigma(r_0) A(r_0)$  and integrate over the ring
to obtain
\be
\int \Sigma(r_0)\Omega
(\Omega_P - \omega_{prec})\ A^2(r_0) dr_0
 - \int \Sigma(r_0)A(r_0)g_{D} dr_0 =  \int \Sigma(r_0)A(r_0)g_{S} dr_0
\label{pr5r2}\ee

In this case we use equation~(\ref{gs21}) to evaluate the second term
of equation~(\ref{pr5r2}), which, after making use of the symmetry
properties of the integral, gives:
\be \int \Sigma(r_0)A(r_0)g_{D} dr_0
= G\ \int_{r_1}^{r_2} \int_{r_1}^{r_2}
\frac{I(q)}{q} \Sigma(r_0')\Sigma(r_0)\ \frac{(A(r_0) -
A(r_0'))^2 }{(r_0 - r_0')^2} dr_0' dr_0\label{gs212}
\ee
which is positive definite. This may also be written 
entirely in terms of $q$ as
\be \int \Sigma(r_0)A(r_0)g_{D} dr_0
= G\ \int_{r_1}^{r_2} \int_{r_1}^{r_2}
I(q)q \Sigma(r_0')\Sigma(r_0)
 dr_0' dr_0.\label{gs213}
\ee
If we now specialize to the case when $q$ is constant
we accordingly write $A(r_0) = A_c + q x$ as  a linear function of radius.
Here $A_c$ is constant and $x = r_0 -r_c$ measures the radial coordinate
relative to the centre of mass of the unperturbed ring
assumed slender. The eccentricity at the ring centre satisfies
is $ e= | A_c| /r_c,$ while the sign of $A_c$
determines whether pericentre is  in the direction
$\phi_0 =0, {\rm or}  \ \ \ \pi$ if negative or positive  respectively.
 This means that by definition
\be \int_{r_1}^{r_2}x\Sigma dx =0 
\label{defx}
,\ee
where the domain of integration for
  the above and similar integrals below is the extent of the ring.
We further adopt a linear form for the precession frequency, thus
\be 
\omega_{prec} = \omega_{prec,0} + \omega'_{prec,0}x ,\ee
where $ \omega_{prec,0}$ and $\omega'_{prec,0}$ represent the precession 
frequency and its derivative evaluated at $x=0.$
 Consistent with the approximations made
here we may also assume the forcing potential term $g_S$ is constant throughout the ring.

The first relation (\ref{pr5r11}) then gives a first relation between
$A_c$ and $q$ in the form
\be
\int \Sigma \Omega
(\Omega_P - \omega_{prec,0})\ A_c dx -\int \Sigma \Omega
\omega'_{prec,0}q x^2 dx 
 =  \int \Sigma g_{S}. dx
\label{pr5r1f}
\ee
The second relation similarly leads to a second which takes the form
$$\int \Sigma \Omega
(\Omega_P - \omega_{prec,0})\ (A_c^2 + q^2 x^2)dx
-2\int \Sigma \Omega\omega'_{prec,0} A_c q x^2 dx
 = $$
\be \int \Sigma\Omega
 \omega'_{prec,0} q^2 x^3 dx
 +G\ \int_{x_1}^{x_2} \int_{x_1}^{x_2}
I(q)q \Sigma(x')\Sigma(x)
 dx'dx
+ \int \Sigma A_c g_{S} dx.
\label{pr5r2f}\ee
We now have two relations which enable the forced
response to be calculated under the assumption
of constant $q.$
These are $A_c,$ whose magnitude, when divided by the radius at the ring centre
gives the eccentricity at the ring centre and $q$ itself
which is the product of the central radius and the eccentricity
gradient. Note that implicit in the thin ring approximation
is the requirement that the magnitude of the eccentricity
gradient significantly exceeds the ratio of the eccentricity to radius.

\section{The relation between $q$ and central eccentricity}
\label{qe}
If we multiply the first relation (\ref{pr5r1f}) by $A_c$ and subtract it from the 
second relation (\ref{pr5r2f}), the terms involving
satellite forcing cancel out and we get a relation between $A_c$
and $q$ which (recalling that $q$ is constant and $\Omega$ is evaluated
in the center of the ring)  takes the form.
$$q\Omega
(\Omega_P - \omega_{prec,0})\int \Sigma
  x^2dx
-\Omega\omega'_{prec,0}A_c\int \Sigma x^2 dx
 = $$
\be q\Omega \omega'_{prec,0} \int \Sigma
x^3dx + G\ \int_{x_1}^{x_2} \int_{x_1}^{x_2}
I(q) \Sigma(x')\Sigma(x)
 dx'dx.
\label{prAqf}\ee

This may be cast in the very simple form

\be A_c =  {q (\Omega_P - \omega_{prec,0})\over \omega'_{prec,0}}
-{q\int
 \Sigma x^3
 dx\over \int \Sigma x^2 dx}
-{GI(q)\left( \int
 \Sigma 
 dx\right)^2\over \Omega\omega'_{prec,0}\int \Sigma x^2 dx} .
\label{Aqfs}\ee

\subsection{Determination of $q$}

We may now use the relation between $q$ and $A_c$ specified above,
to eliminate $A_c$ in the first relation (\ref{pr5r1f}) and so obtain 
$q$ in terms of the satellite forcing. This gives

$$
 q\left(
 { (\Omega_P - \omega_{prec,0})^2\over (\omega'_{prec,0})^2}
-{(\Omega_P - \omega_{prec,0})\over \omega'_{prec,0}}{\int x^3\Sigma
 dx\over \int x^2\Sigma dx}
-{\int \Sigma
x^2 dx\over \int \Sigma dx}\right)$$
\be
-{(\Omega_P - \omega_{prec,0})GI(q)\left( \int
 \Sigma
 dx\right)^2\over \Omega(\omega'_{prec,0})^2\int \Sigma x^2 dx}
 = {g_{S}\over \Omega \omega'_{prec,0}} . 
\label{qSf}
\ee

Thus equations~(\ref{Aqfs}) and~(\ref{qSf}) give the response parameters
$A_c$ and $q$ directly in terms of the external forcing.

\subsection{The linear regime}
 When the response is in the linear regime $q$ is small
and $I(q)= q/2.$ In this case   the right hand side of
(\ref{qSf}) is proportional to $q$  and we have
\be a_1q 
= {g_{S}\over \Omega \omega'_{prec,0}},
\label{qcub}
\ee
where

$$
  a_1 = { (\Omega_P - \omega_{prec,0})^2\over (\omega'_{prec,0})^2}
-{\int \Sigma
x^2 dx\over \int \Sigma dx}
-{(\Omega_P - \omega_{prec,0})\over \omega'_{prec,0}}{\int x^3\Sigma
 dx\over \int x^2 \Sigma dx}$$
\be -{(\Omega_P - \omega_{prec,0})G\left( \int
 \Sigma
 dx\right)^2\over 2 \Omega(\omega'_{prec,0})^2\int \Sigma x^2 dx}
.\ee

The response is singular when $a_1=0$. Regarding this as an equation for
$\Omega_P - \omega_{prec,0},$ we have a quadratic with two real roots
indicating a singular response for certain pattern speeds. However, the
relation (\ref{qSf}) is in fact a nonlinear one (through the functional form
of $I(q)$) to determine $q$ and the nonlinearity present can remove such
singularities. This becomes apparent when we expand to the next highest
order in $q.$ Then one obtains a cubic equation for $q$ that can always be
solved because the coefficients of $q^3$ and $q$ never vanish
simultaneously. This cubic takes the form \be a_1q - q^3{3(\Omega_P -
\omega_{prec,0})G\left( \int \Sigma dx\right)^2\over
8\Omega(\omega'_{prec,0})^2\int \Sigma x^2 dx} = {g_{S}\over \Omega
\omega'_{prec,0}} . \label{qcub2} \ee

\noindent However, as indicated above we must be cautious
about using the above determinations of $A_c$ and $q$ when the 
eccentricity response is large because then the assumption
of constant $q$ and the neglect of the dependence
of externally induced  orbital precession on the eccentricity
may no longer be valid.

\section{Application to the Titan ringlet}
\label{Titan}

We here consider the Titan ringlet for which the ring precession rate is
close to the orbital frequency of the satellite Titan. This ringlet is
therefore a candidate for having its eccentricity forced by Titan. We now
consider the forcing potential.

\subsection{The satellite potential}
For the low frequency $m=1$ forcing considered here
the perturbing effect of the satellite arises through
\be g_{S} = - \frac{1}{2\pi}
\int_0^{2\pi} \left( \cos(\phi_0) {\partial \psi_{s} \over \partial
r} - 2\sin (\phi_0){1\over r} {\partial \psi_{s} \over \partial \phi_0}\right)
 d\phi_0
\ee

\noindent For the satellite we neglect the orbital eccentricity
and expand $\psi_{s}$ to leading order in $r/a_S,$ $a_S$
being the semi-major axis of the satellite orbit.
For the forcing considered here, recalling that in the frame rotating with the
orbital frequency $\Omega_p,$  
only secular terms $\propto \cos(\phi_0)$  are significant,
we obtain, including the indirect potential,  to leading order
\be
\psi_{s} = - {3 GM_S r^3\over 8 a_s^4}\cos(\phi_0).
\label{phis}
\ee
Then  evaluating at the ring center $r=r_c \equiv a,$ we obtain
\be g_S =  {15 GM_S r_c^2\over 16 a_s^4}. \ee 

\noindent We are now ready to apply equations~(\ref{Aqfs}) and~(\ref{qSf}) to the Titan ringlet.

\subsection{Parameterizing the dynamical model}

We define the dimensionless parameter $\eta$ through \be \Omega_P -
\omega_{prec,0}= \eta\ \omega_{prec,0}'\ \Delta a.  \label{eta} \ee This
defines the resonance where the precession rate induced by the planet and
Titan match in the approximation that the former can be represented by a
first order Taylor expansion about the ring centre. It gives the resonance
location at a distance $\eta \Delta a$ from the ring centre. 

We also find it convenient to define the dimensionless quantities $\gamma_n$
related to the ring surface density profile through 
\be \gamma_n = {\int
x^n\Sigma dx \over (\Delta a)^n\int \Sigma dx}.\ee 
The location of the center of mass of the ring,$r_c$, is the origin of the
coordinate $x$ (equation~(\ref{defx})). Then, inside the ring  we have
$|x| < \Delta a$ --the extreme  case is when the
 mass is concentrated  at  a ring edge. Thus, it is verified that $|\gamma_n|
< 1$. We also notice that $\gamma_n > 0$ when $n$ is even and, when
$n$ is odd, $\gamma_n = 0$ if the surface density is symmetric.  

Further we introduce the parameter $\Phi$ which measures the importance 
of self-gravity with respect to differential precession:
\be \Phi = {G {\overline \Sigma}\over 2\Omega \omega_{prec,0}'(\Delta a)^2}, 
\label{PHI}
\ee
where the mean surface density is defined as: 
\be   {\overline \Sigma} = {\int \Sigma dx \over \Delta a}, \ee
notice that since the precession frequency decreases with distance 
(i.e. $\omega_{prec,0}' < 0$), then $\Phi$ is defined as a negative quantity. 

\noindent In terms of these dimensionless parameterizations,
equations~(\ref{Aqfs}) and~(\ref{qSf}) can be reduced to
\be A_c  =q \Delta a\ a_2, \label{F1}
\ee
where 
\be
a_2 = \eta- {\gamma_3\over \gamma_2} - {\Phi \over \gamma_2}, 
\label{a2}
\ee
and
\be q\Delta a = {g_S\over \Omega\omega_{prec,0}'\Delta a}
 \left( \eta\ a_2  - \gamma_2 \right)^{-1}. \label{F2}\ee

\subsection{Estimation of the surface density and the location of the secular
resonance}

The presently available observations of the Titan ringlet do not enalble one
to determine all its dynamical and physical parameters. Optical depth
profiles give the most accurate description of the ring, thus the surface
density can only be inferred by making assumptions on the physical properties
of the ring particles or through dynamical models which, like the present
one, rely on various assumptions (see for example Goldreich and Tremaine
1979). Moreover, there are considerable uncertainties in the values of the
multipole moments of Saturn (for up-to-date values see the JPL-Solar
System Dynamics website: {\it http://ssd.jpl.nasa.gov/sat\_gravity.html}).
In fact, the uncertainties in the precession frequency of the ring due to
the uncertainties in multipole  moments of Saturn, imply an error in the
location of the secular resonance that is of the order of the width of the
ring.  

Using our model we shall attempt to put some constraints on
the value of $\Phi$ and so, on the mean surface density , ${\overline \Sigma}$,
as well as on the resonance location parameter, $\eta$, for the Titan ringlet.

\noindent We shall adopt the following values:
$ a = 77871 km, \Delta a = 25 km,
 e = 2.6 \times 10^{-4}, \delta e = (1.4 \pm .4) \times 10^{-4},$ (Porco et al 1984).
Thus $q =a\delta e/\Delta a \approx 0.44 \pm 0.18$. 

\noi The closest approach between the satellite and the ring occurs at apoapse
(Porco et al. 1984). Then, according to the definitions of the satellite
potential (equation~(\ref{phis})) and of the radial displacement
(equation~(\ref{xi_r})) adopted here, when $\phi=0,$  $\xi_r > 0$. Thus, for
the Titan ringlet   we have $A_r(r_0) >0$ as well as $A_c > 0$. 


Now, we shall rewrite equations~(\ref{F1}) and~(\ref{F2}) as:
\bea
a_3 &=& \eta\ a_2  - \gamma_2 \label{a3eq} \\
-\Phi &=& (a_2 - \eta)\ \gamma_2 + \gamma_3, \label{mphi}
\eea
where:
\be
a_3 = {g_S\over \Omega \omega_{prec,0}' q (\Delta a)^2}.
\label{a3}
\ee
Notice that $a_3<0$ since $\omega_{prec,0}'<0$ and $g_S>0$.

We can eliminate $\eta$ from equations~(\ref{a3eq}) and~(\ref{mphi}), 
to obtain a quadratic expression of $\Phi$ as a function of the form factor $\gamma_2$:
\be
-\Phi = -\frac{1}{a_2}\ \gamma_2^2 + (a_2 - \frac{a_3}{a_2})\ \gamma_2 +  \gamma_3.
\label{cuadphi}
\ee

Similarly, one can eliminate $\gamma_2$ and express $\Phi$ as a function of $\eta$:  
\be
-\Phi = - a_2\ \eta^2  + (a_2^2 + a_3)\ \eta - a_3 a_2 + \gamma_3.
\label{cuadphi2}
\ee
One must recall that $\eta$ and $\gamma_2$ are related by
equation~(\ref{a3eq}), which acts as a constraint for $\eta$ since it must be verified
that $\gamma_2 < 1$.


Note also that $a_2$ can be   expressed as a {\it relative eccentricity}, 
\[a_2 \approx e_c/ \delta e\] through the use of equation~(\ref{F1}). 
The quadratic terms in equations~(\ref{cuadphi}) and~(\ref{cuadphi2}) depend
on $a_2$. Then, if the relative eccentricity is large, is:
$-\Phi \approx a_2 \gamma_2 + \gamma_3$. Whereas if $a_2$ is small, $ -\Phi
\approx a_3\ \eta - a_3 a_2 + \gamma_3$.

One may expect to find observed values of $\gamma_2$ between those of two critical
cases. One extreme case for the shape of the surface-density distribution is when
it is constant throughout the ring, 
\[\Sigma^{(1)} \neq \Sigma^{(1)}(x),\] 
the other when the whole mass is
concentrated at both edges, as:
\[\Sigma^{(2)} \propto \delta(x - \Delta a/2) + \delta(x + \Delta a/2)
, \] were $\delta$ is the Dirac-$\delta$
function. In these cases the values of $\gamma_2$ are
given by $\gamma_2^{(1)} = 1/12$ and $\gamma_2^{(2)} = 1/4$ respectively.   

We can estimate the values of $a_2$ and $a_3$ for the Titan ringlet as: \bea
a_2 &\approx& \left(\frac{e}{\delta e} \right)_{Obs} \approx 1.86
\label{a22} \\ a_3 &\approx& - 1.38, \label{a32} \eea where these estimates
are obtained from the observed parameters previously quoted and, in
equation~(\ref{a32}), the precession frequency gradient, $\omega_{prec,0}'$,
is computed using the $J_2$, $J_4$ and $J_6$ cofficients associated with the
multipole moments of Saturn
given in the JPL-SSD website: $J_2=\ (16292 \pm 7)\times10^{-6}$, $J_4=\
(931 \pm 32)\times10^{-6}$, $J_6=\ (91 \pm 32)\times10^{-6}$ (Jacobson
2004). We computed the precession rate of the ring due to the oblateness of
Saturn and the secular perturbation by Titan \footnote{We use an expansion
up to $(R_{Sat}/a)^6$ in the oblateness term. For the secular term, we used
an approximation up to first order in $e$ for the perturbing function (see
for example Murray \& Dermott 1999), where we have used a value of
$a_{Titan}=\ 1221900.\ km$ for the semimajor axis of the satellite and of
$m_{Titan} = 1345.5 \times 10^{23} gr$, for its mass.} and we obtain a value
of $a_{Res}=\ (77846.\ \pm\ 11.)\ km $, for the location of the $1:0$
secular resonance, where the errors only consider the uncertainties in the
values of Saturn's $J$-coefficients.  

For the symmetric case, we have plotted the relationships set by
equations~(\ref{a3eq}), (\ref{cuadphi}) and~(\ref{cuadphi2}) for the values
of $a_2$ and $a_3$ corresponding to the Titan ringlet. 

We may recall that, if the surface density is symmetric with respect to the
center of mass, then $\gamma_3 = 0$ and that 
inside the ring  $|x| \leq \Delta a/2$ and $|\eta| \leq 0.5$.

Figure~(\ref{fig1}) shows ${\overline \Sigma}$ vs. $\gamma_2$ --where
${\overline \Sigma}$ is related to $\Phi$ by equation~(\ref{PHI}).  It can
be seen that the scale of the mean surface density is  tens of grammes
per cubic centimetre, in particular when $ \gamma_2^{(1)} \leq \gamma_2 \leq
\gamma_2^{(2)}$. In the symmetric case, the agreement with the solution
quoted in Porco et al. (1984) occurs at a value of $\gamma_2 \approx
\gamma_2^{(1)}$,
however, in that case the numerical values of various parameters are different. 

In figure~(\ref{fig2}) we plot $\eta$ vs. $\gamma_2$.  This figure gives,
for a symmetric profile, the location of the secular resonance as a function
of the form of the surface density, such that the ring precesses rigidly
when the effects of inter-particle collisions are negligible. The allowed
values of $\eta$, set by equation~(\ref{a3eq}) and $\gamma_2<1$, imply that,
given the condition of solid precession, the location of the secular
resonance lies outside the ring ($|\eta| > 0.5$) for the symmetric case with
$ \gamma_2^{(1)} \leq \gamma_2 \leq \gamma_2^{(2)}$. We may recall that if
$\eta \sim 0$, then the assumptions made to obtain these relationships break
down, because the eccentricities resonantly-excited may become very large.
Thus, the rather large values of $\eta$ obtained assure us that the first
order approximation adopted here is consistent with the case. However, as
mentioned earlier, the uncertainties involved are large.

Finally, in figure~(\ref{fig3}) we plot ${\overline \Sigma}$ as a function
of $\eta$. This figure presents the required values of the mean
surface-density to produce the balance as a function of the location of the
secular resonance. Naturally, the result reflects the particular choice of
parameters adopted. All the solutions obtained require $\eta < 0$, hence the
the center of mass of the ring must be located {\it interior} (i.e. towards
the satellite) with respect to the secular resonance (see
equation~(\ref{eta})). 

If the surface density profile were not to be symmetric then these estimates
can be somewhat altered. By definition, the magnitude of
$\gamma_3$ is bounded by that of $\gamma_2$ as: 
\[|\gamma_3| \leq
\frac{|x_{max}|}{\Delta a}\ \gamma_2 \leq \gamma_2. \]
In figure~(\ref{fig1}) we also show ${\overline \Sigma}$ as a function of
$\gamma_2$, for the extreme cases in which $|\gamma_3|=\gamma_2$.   

On the other hand, we do not expect the effect of the collisional terms to be
negligible, as assumed in these calculations. We will explore their exact 
significance in a forthcoming paper. We will discuss in the next section how
physical particle interactions could 
affect the results obtained here. 

%
\section{Discussion}
\label{discu}

















       
In this paper we have used a model of a thin slender self-gravitating ring
in orbit about a dominant central mass to produce relationships between
its physical  parameters. These relationships can then be applied to
real systems to make predictions about their physical and/or dynamical
properties. 

We view the ring in uniform precession as sustaining a global
non-axisymmetric $m=1$ mode of oscillation. We have considered particularly
the case in which the precession frequency is in a secular resonance with
the orbital frequency of an external satellite, which is the case of the
Titan ringlet of Saturn. 

A condition for the ring to be able to maintain a $m=1$ mode with single
slow pattern speed is obtained. It can be expressed as an integral condition
for the pattern or normal mode to precess at a uniform rate
(Eq.~(\ref{pr4})) that requires the correct balance between differential
precession, self-gravity, secular-resonant forcing and collisional effects. 
From this condition we have obtained two relationships
(equations~(\ref{pr5r11}) and~(\ref{pr5r2})) which combine different observable
parameters of the ringlets.

We argued that the scale of the perturbations is of the order of the width
of the ring, because we are interested in cases in which the differential
precession is compensated mainly by self gravity and the eccentricity is
small. Hence, a first order expansion is a good approximation and so, we 
take the eccentricity gradient as constant, which, indeed, is the case in real
systems (Graps et al. 1995).

In the linear case, the two expressions obtained give simple relationships
(equations~(\ref{pr5r1f}) and~(\ref{pr5r2f})) between the eccentricity at
the center, $e_c$, the eccentricity gradient, $q$, the location of the
secular resonance, $\eta$ and the mean value of the  surface density
${\overline \Sigma}$, and its form, expressed by the 
$\gamma_n$'s.  The relationships can be further simplified by assuming that
the system is also linear in $q$ (equations~(\ref{Aqfs}) and~(\ref{qcub})).  

Finally we have applied this model to the Titan ringlet. The scale of the
surface density obtained is consistent with previous estimations (Porco et
al. 1985), set at the order of $\sim 10\ gr\ cm^{-2}$. Our symmetric solutions
imply that the secular resonance is outside the ring (see
figure~(\ref{fig3})). However, the distance between the ring and the
location of the secular resonance obtained is smaller than the total
uncertainty in the location of the later, given the errors in the
determination of the multipole moments of Saturn.  

Moreover, additional physics may need to be considered. The fact that
all the observed systems have shown a positive value of $q$ is an indication
that the differential precession is mainly balanced by self-gravity
(Borderies et al 1983, Papaloizou and Melita 2004). However, it is also very
likely that physical collisions play an important role in setting up the
rigid precession, particularly at the edges, where the relative motion
is larger (see for example Chiang and Goldreich 2001).  It is also believed
that the real systems are close-packed at their narrowest point --as it
seems to be indicated by the observations of the Uranian rings (see for
example French et al. 1984). The narrowest points of the observed eccentric
ringlets occur at periapse, since $q>0$. Thus, the collisional contribution
can be estimated to be impulsive, i.e. entirely concentrated at the {\it
pinch}. At periapse, the geometry of the collisions at the edges is such
that the collisionally produced orbital  phase-shift produced {\it reinforces} the
differential-precession induced by the oblateness of the central planet (see
for example Moulton 1935). Thus, if a considerable collisional contribution
arises mainly at the pinch, self-gravity must counteract a greater
differential precession. If that is the case, the values of the masses of
the real eccentric ringlets will be {\it larger} than the estimates produced
here. In fact, the masses of eccentric ringlets estimated from observations
have always turned out to be greater than the theoretical estimates obtained
with models that consider only the self-gravity of the ringlet (see for
example Mosqueira \& Estrada 2001). It is straightforward  to
include collisional effects when they are assumed to act impulsively at pericentre. 
 --more terms are to be retained
when considering the two conditions for rigid precession given by 
(equations~(\ref{pr5r1}) and~(\ref{pr5r2})). We can gain some knowledge of
these collisional effects if an observed profile is
available --from which we can extract density estimates. We will present a
more detailed analysis of some known eccentric ringlets in a forthcoming
article.

%
%

\section*{References}

\noi Borderies, N., Goldreich, P. and S. Tremaine. 1983. The dynamics
of elliptical rings. {\it Astron. J.} {\bf 88}, 1560-1568.

\noi Chiang,E.I. and P.Goldreich,P. 2000. Apse Alignment of Narrow
Eccentric Planetary Rings. 
{\it Astrophys. J.} {\bf 540}, 2, 1084-1090.

\noi Dermott, S.F. and C.D. Murray. 1980. The origin of the
eccentricity gradient and the apse alignment of the $epsilon$-ring of
Uranus. {\it Icarus} {\bf 43}, 338-349.

\noi French R.G., Nicholson P.D., Porco, C and Marrouf, E.A. 1984. Dynamics
and Structure of the Uranian rings. In,
Planetary Rings, Richard Greenberg \& Andre Brahic,
eds., University of Arizona press, Tucson, Arizona, 513-561.

\noi Goldreich, P. and C.C. Porco. 1987. Shepherding of the Uranian
rings. II. Dynamics. {\it Astron. J.} {\bf 93}, 730-737.


\noi Goldreich, P. and S. Tremaine. 1979. Precession of the epsilon
ring of Uranus.  {\it Astron. J.} {\bf 84}, 1638-1641.

\noi Goldreich, P. and S. Tremaine. 1981. The origin of the
eccentricities of the rings of Uranus. {\it Astrophys. J.} 
{\bf 243} 1062-1075. 


\noi Graps, A. L., M.R. Showalter, J.J. Lissauer and
D.M. Kary. 1995. Optical Depths Profiles and Streamlines of the
Uranian (epsilon) Ring. {\it Astron. J}, {\bf 109}, 2262-2273.

\noi Jacobson, R.A. 2004. The orbits of the major Saturnian satellites and the
gravity field of Saturn from spacecraft and Earth based observations.
{\it Submitted to the Astronomical Journal}.

\noi Longaretti, P.Y. and Rappaport, N. 1995.
Viscous overstabilities in dense narrow planetary rings.
 {\it Icarus}, {\bf 116}, 376-396.



\noi Murray C.D. and Dermott S. 1999. Solar System dynamics. Cambridge
University press. Cambridge, United Kingdom.

\noi Papaloizou J.C.B. and Melita M.D. 2004. Structuring narrow eccentric
planetary rings. {\it Icarus}. Submitted.




\noi Mosqueira, I.and P.R. Estrada. 2002. Apse Alignment of the Uranian
Rings. {\it Icarus}, {\bf 158}, 2, 545-556.

\noi Moulton, F.R. 1935.
An introduction to celestial mechanics. Ed: The Macmillan company, London.

\noi Porco C., Nicholson P.D., Borderies, N., Danielson G.E., Goldreich P.,
Holberg J.B. and Lane A.L. 1984. The eccentric Saturnian rings at 1.29R$_S$
and 1.45R$_S$. {\it Icarus}. {\bf 60}. 1-16. 

\noi Shu F.H., Yuan C. and J.J. Lissauer. 1985. Nonlinear spiral
density waves: An inviscid theory. {\it Astrophys. J.} {\bf 291},
356-376

\noi Tyler, G.L., V.R.  Eshleman, D.P. Hinson, E.A. Marouf,
R.A. Simpson, D.N. Sweetnam, J.D. Anderson, J.K. Campbell, G.S. Levy
and G.F. Lindal.  1986. Science, 233, 79-84.  Voyager 2 radio science
observations of the Uranian system Atmosphere, rings, and satellites.
{\it Science}, {\bf 233}, 79-84.

\newpage

\begin{figure}[ht]
\includegraphics[width=110mm,height=70mm,angle=0]{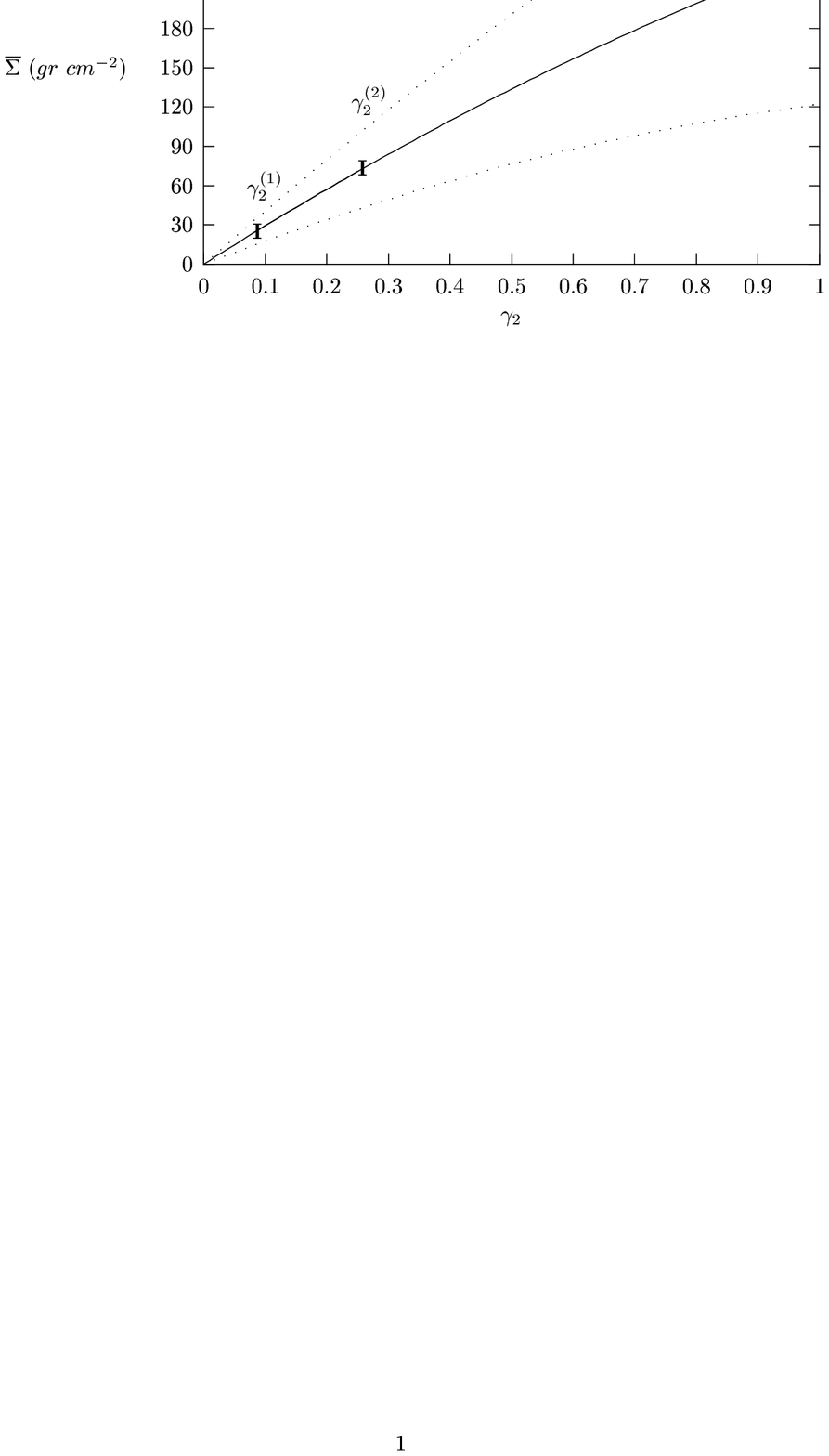}
\caption{The estimated mean surface density, ${\overline \Sigma}$, as a function of
$\gamma_2$ for a symmetric profile ($\gamma_3=0$). Notice the value of
${\overline \Sigma}$ when $\gamma_2$ is between the critical values 
$\gamma_2^{(1)} = 1/12$ and $\gamma_2^{(2)} = 1/4$ (indicated by the markers).
 The dotted lines correspond to the non-symmetric case where $|\gamma_3| =
\gamma2$.}
\label{fig1}
\end{figure}
                           
\newpage
             
\begin{figure}[ht]
\includegraphics[width=110mm,height=80mm,angle=0]{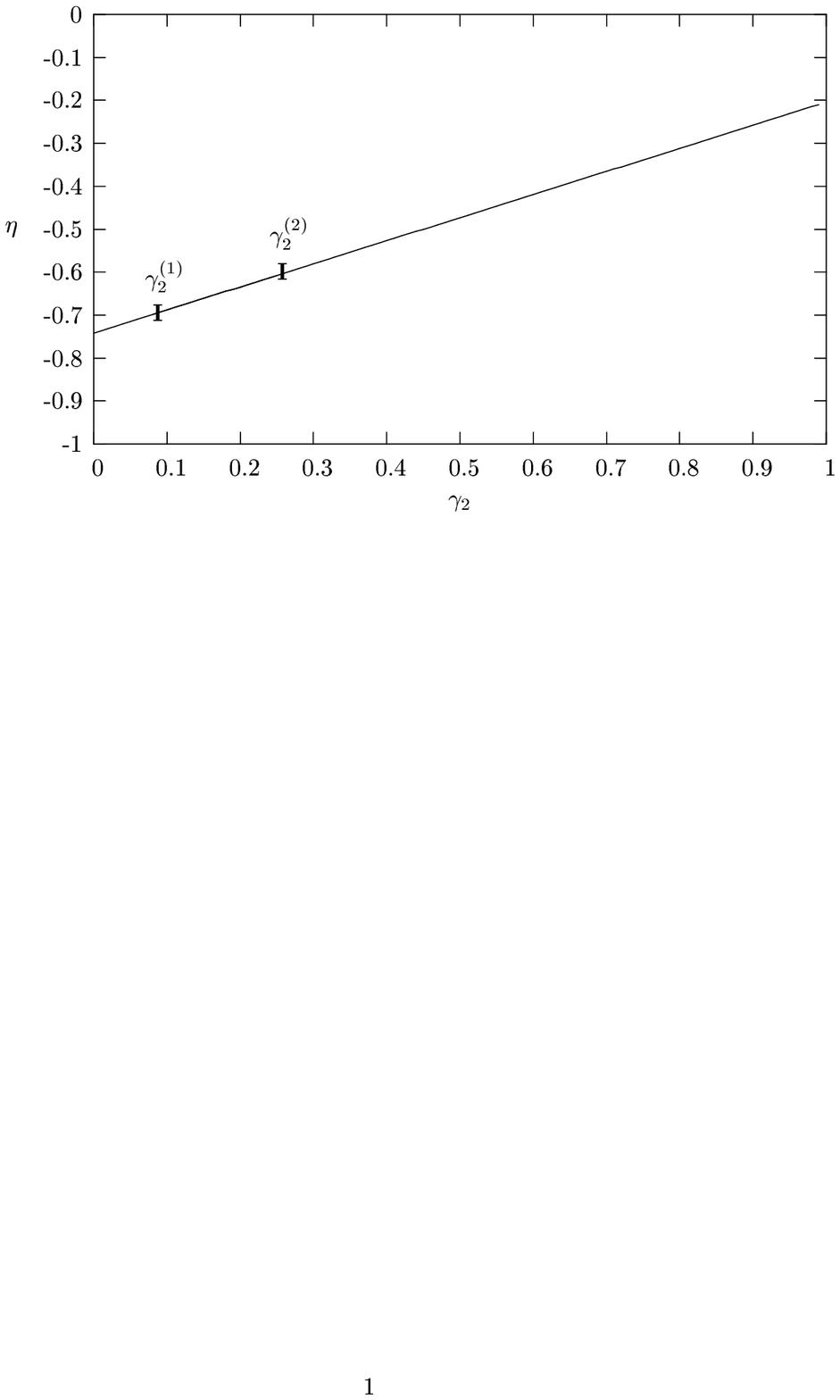}
\caption{The location of the resonance $\eta$, as a function of
$\gamma_2$ ($\gamma_3=0$) that gives the correct balance enabling the rigid precession of
the ring, when the collisional terms are neglected. Notice that solutions
with 
$\gamma^{(1)}
\le
\gamma \le \gamma^{(2)}$, 
correspond to resonance locations outside the ring ($|\eta| > 0.5$).}
\label{fig2}
\end{figure}

\newpage
             
\begin{figure}[ht]
\includegraphics[width=110mm,height=60mm,angle=0]{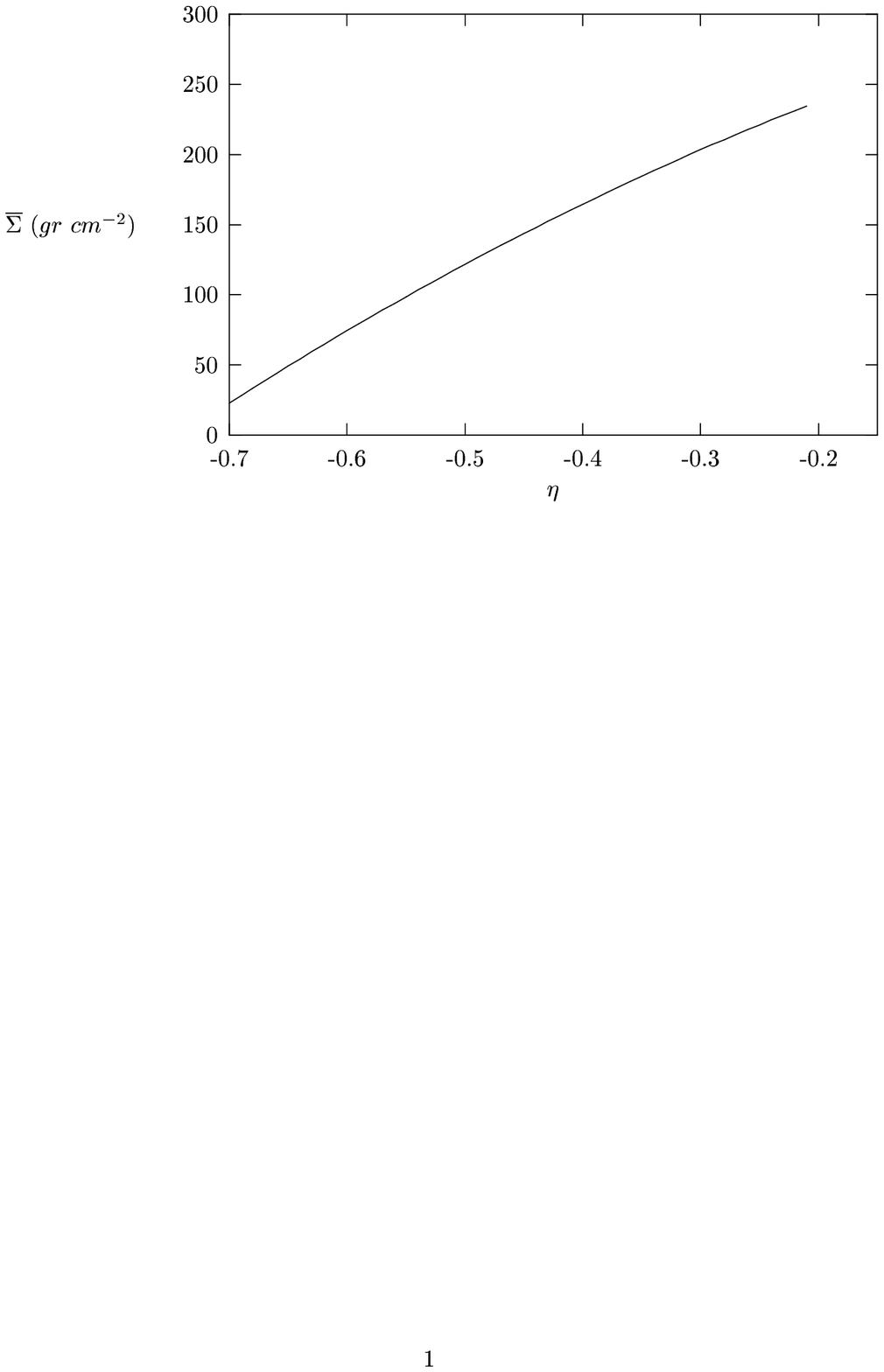}
\caption{The estimated mean surface density, ${\overline
\Sigma}$, as a function of
$\eta $ ($\gamma_3=0$). A negative value of $\eta$ implies that
$\Omega_P > \omega_{prec,0}$ (equation~(\ref{eta})), i.e. the
center of mass of the ring is exterior -from the planet- to the location of secular resonance.}
\label{fig3}
\end{figure}

\end{document}